\documentclass[runningheads,a4paper]{llncs}
\usepackage{amssymb}
\usepackage{amsmath}
\usepackage{hyperref}

\usepackage{graphicx}
\usepackage{url}
\newcommand{\keywords}[1]{\par\addvspace\baselineskip
\noindent\keywordname\enspace\ignorespaces#1}
\begin{document}

\mainmatter

\title{Numerical integration in arbitrary-precision \\ ball arithmetic}  
\titlerunning{Numerical integration in ball arithmetic} 
\author{Fredrik Johansson}
\authorrunning{Johansson}
\institute{
LFANT -- INRIA -- IMB, Bordeaux, France\\
\email{fredrik.johansson@gmail.com}\\ 
\texttt{http://fredrikj.net}
}
\maketitle

\begin{abstract}
We present an implementation of arbitrary-precision numerical integration
with rigorous error bounds in the Arb library.
Rapid convergence is ensured for piecewise complex analytic integrals
by use of the Petras algorithm, which combines
adaptive bisection with adaptive
Gaussian quadrature where error bounds are determined via
complex magnitudes without evaluating derivatives.
The code is general, easy to use, and efficient,
often outperforming existing non-rigorous software.

\keywords{Numerical integration, interval arithmetic, special functions}
\end{abstract}

\section{Introduction}

Many users can attest that there is a non-negligible chance of getting an
incorrect answer
when asking a numerical package or computer algebra system
for an approximation of a definite integral
$\smash{\int_a^b f(x) dx}$,
as rapid variation, narrow peaks, non-smooth points,
cancellation or ill-conditioned numerical evaluation of $f$
are prone to break widely used heuristic numerical integration methods.

One remedy is to compute rigorous error bounds using interval arithmetic.
However, little work has been done to date on efficient arbitrary-precision implementations.
Here, we present a new implementation of rigorous numerical
integration in Arb,
a C library for ball arithmetic\footnote{Arb (\url{http://arblib.org}) is open source (GNU LGPL) software.
For documentation and example code related to this paper, see \url{http://arblib.org/acb_calc.html}.}~\cite{Johansson2017arb}.
The integration code is easy to use directly in C, or can be wrapped
from high-level languages.
For example, an interface in Sage~\cite{sage} exists (thanks to Marc Mezzarobba
and Vincent Delecroix), which we demonstrate by computing $\smash{\int_0^8 \, \sin(x+e^x) \, dx}$:

\begin{small}
\begin{verbatim}
sage: C = ComplexBallField(333)    # 333-bit precision
sage: C.integral(lambda x, d: sin(x+exp(x)), 0, 8)
[0.347400172657247807879512159119893124657456254866180183885492713616748
21398878532052968510434660 +/- 5.97e-96]
\end{verbatim}
\end{small}

We obtain nearly 100 digits with a rigorous error bound
in 0.04 seconds (0.02~s when using C directly).
This relatively difficult test integral ($f(x)$ changes sign 950 times)
was introduced by Rump~\cite{rump2010verification} who observed
that the \texttt{quad} function in Matlab took over a second only to return the erroneous \texttt{0.2511}
(Rump's interval package Intlab computes 7 digits in about one second; see also \cite{mahboubi2016formally}).

\section{Algorithm and implementation}

\vspace{-1.3mm}
We consider integration of a function $f : \mathbb{C} \to \mathbb{C}$
on a segment $[a,b]$, $a,b \in \mathbb{C}$.
We represent real numbers as mid-rad intervals (balls) $[m \pm r]$
and complex numbers as rectangles $[m_1 \pm r_1] + [m_2 \pm r_2] i$
(which we also refer to as balls with slight abuse of terminology).
True complex balls $B(m_1+m_2 i,r)$ would sometimes provide slightly better bounds,
but rectangles are usually more convenient.

The user supplies the integrand $f$ as a pointer to a C
function \texttt{func} implementing its evaluation (we refer to the documentation for the detailed API).
In effect, \texttt{func} gets called with the
argument $z$ and an extra flag $d$.
If $d = 0$, \texttt{func} is to evaluate $f(z)$ without any assumptions
about regularity.
If $d = 1$, \texttt{func} is to evaluate $f(z)$ and also check
that $f$ is analytic on $z$, returning a non-finite ball (e.g.\ NaN) otherwise.
For meromorphic $f$, the user can ignore $d$ since~$f(z)$
automatically blows up at poles, but $d$ needs to
be handled for functions with branch cuts like $\sqrt{z}$ and $\log(z)$
(here by checking whether $z$ overlaps $(-\infty,0]$).

We use the Petras algorithm~\cite{petras2002self}, which combines
bisection with Gaussian quadrature of variable degree $n$.
Error bounds for Gaussian quadrature use complex magnitudes.
If $f$ is analytic with $|f| \le M$ on an ellipse $E$ with foci $\pm 1$ and
semiaxes $X,Y\!\!$,
then $\smash{|\!\int_{-1}^1 \!f(x) dx - \sum_{k=1}^n \!w_k f(x_k)| \le M \rho^{-2n} C_{\rho}}$, $\rho = X + Y$,
where e.g.\, $C_{\rho} < 50$ if $\rho > 1.1$.
The tradeoff
is that a larger $E$ increases~$M$, with $M = \infty$ if $E$ hits a singularity of $f$,
but also improves convergence as $n \to \infty$.
Of course, the computed bound for $M$ will not just depend on the function~$f$
but also on the stability of its evaluation in ball arithmetic if $E$ is large.

Degree adaptivity ensures near-optimal complexity ($O(p)$ evaluations of $f$)
for analytic $f$ at high precision $p$, while
space adaptivity (bisection) helps if there are
singularities near $[a,b]$ or if the ball enclosures are not optimal.
For piecewise analytic $f$ with discontinuities on $[a,b]$ the complexity is typically $O(p^2)$,
i.e.\ a bit worse but still polynomial in $p$.
Degree or space adaptivity used alone would
give $2^{O(p)}$ complexity or fail to converge
for common types of integrals.


Our version of the integration algorithm can be described as follows:
\vspace{-2mm}
\begin{itemize}
\item Initialize sum $S \gets 0$, subinterval work queue $Q \gets [(a,b)]$.
\item While $Q = [(a_1,b_1), \ldots, (a_N,b_N)]$ is not empty:
\begin{enumerate}
\item Pop $(\alpha,\beta) = (a_N,b_N)$ from $Q$.
\item Compute the direct box enclosure $I = (\beta-\alpha) f([\alpha,\beta])$ (evaluating $f$ on $z = [\alpha,\beta]$ with $d = 0$).
      If $I$ meets the tolerance goal, if $\alpha,\beta$ overlap, or if evaluation limits have been exceeded, set $S \gets S + I$ and go to 1.
\item Try to find an ellipse $E$ with foci $(\alpha,\beta)$ and an $n \le n_{\text{max}}$
such that $f$ is analytic on $E$ (evaluating $f(E)$ with $d = 1$) and the error bound for $n$-point Gaussian quadrature determined via $|f(E)|$
meets the tolerance goal. If successful, compute this integral $J$, set $S \gets S + J$ and go to 1.
\item Interval bisection: let $\smash{m = \tfrac{\alpha+\beta}{2}}$ and extend $Q$ with $(\alpha,m)$, $(m,\beta)$.
\end{enumerate}
\end{itemize}
\vspace{-1.5mm}

Compared to Petras~\cite{petras2002self}, there are minor differences.
Our $\rho$ is not fixed; we
try several sizes of $E$ in step 3 to reduce $n$.
The handling of tolerances is slightly different.
We also compute quadrature nodes $(w_k,x_k)$ at runtime, without using
pre-made tables.
A key point is that generating nodes
for high-precision Gaussian quadrature
used to be considered too costly~\cite{bailey2011high}, but
the recent work~\cite{JohanssonMezzarobba2018fast} solves this problem.\footnote{Clenshaw-Curtis 
or double exponential quadrature could be used instead of Gaussian quadrature, but
typically require more points for equivalent accuracy.
We could also use Taylor series, but
this makes supplying $f$ more cumbersome for the user,
and computing $f,f'\ldots,f^{(n)}$ tends to be more costly than $n$
evaluations of $f$.}
With default settings, computing nodes takes a few milliseconds for 100-digit precision
and a few seconds for 1000 digits.\footnote{In benchmark results, we omit the first-time nodes precomputation overhead.}
Nodes are automatically cached, so this cost is amortized
for repeated integrations at the same or lower precision
(possible~$n$ are restricted to a sparse sequence $\approx 2^{k/2}$
to avoid computing nodes for many nearby $n$).
As an optional tuning parameter,
the user can change the allowed range of $n$ which defaults
to $n_{\text{max}} = 0.5 p + 60$. 
\vspace{-2mm}

\subsection{Tolerances and evaluation limits}

Besides the working precision $p$, the user specifies
absolute and relative tolerances $\varepsilon_{\text{abs}}$
and $\varepsilon_{\text{rel}}$.
In effect, the algorithm attempts
to achieve an error of $\max(\varepsilon_{\text{abs}}, V \varepsilon_{\text{rel}})$
where $V$ is the magnitude of the integral.
Reasonable values (used as defaults by the Sage wrapper)
are $\varepsilon_{\text{abs}} = \varepsilon_{\text{rel}} = 2^{-p}$.
Other values can be useful, e.g.\ if
low accuracy is sufficient but a higher $p$ must be used for numerical reasons.
One might also set $\varepsilon_{\text{abs}} = 0$
to use relative tolerance only,
though for efficiency, it is better to supply
$\varepsilon_{\text{abs}} \approx V \varepsilon_{\text{rel}}$ if an estimate for $V$ is known
when $V \not\approx 1$. This Sage code shows
computation of $\int_0^1 e^{-1000+x} \sin(10x) dx$:

\vspace{-1.5mm}
\begin{small}
\begin{verbatim}
sage: C = ComplexBallField(64); f = lambda x, _: exp(-1000+x)*sin(10*x)
sage: C.integral(f, 0, 1)
[+/- 4.09e-434]                                      # time 0.013 ms
sage: C.integral(f, 0, 1, abs_tol=0)
[1.574528586972758e-435 +/- 7.36e-451]               # time 1.1 ms
sage: C.integral(f, 0, 1, abs_tol=exp(-1000)/2^64)
[1.574528586972758e-435 +/- 7.27e-451]               # time 0.38 ms
\end{verbatim}
\end{small}
\vspace{-3mm}

Conversely, for a large integrand:

\vspace{-3mm}
\begin{small}
\begin{verbatim}
sage: f = lambda x, _: exp(1000+x)*sin(10*x)
sage: C.integral(f, 0, 1)
[6.11102916709322e+433 +/- 1.98e+418]                # time 1.1 ms
sage: C.integral(f, 0, 1, abs_tol=exp(1000)/2^64)
[6.11102916709322e+433 +/- 1.95e+418]                # time 0.39 ms
\end{verbatim}
\end{small}
\vspace{-2.1mm}

In reality, $\varepsilon_{\text{abs}}$ and $\varepsilon_{\text{rel}}$ are only \emph{guidelines}
and the algorithm does not strictly
achieve the goal $\max(\varepsilon_{\text{abs}}, V \varepsilon_{\text{rel}})$.
Indeed, due to the fixed working precision and possibly inexact parameters,
the goal cannot generally be achieved.
It is implied that the user will work with some guard bits and
if needed adjust ($p$, $\varepsilon_{\text{abs}}$, $\varepsilon_{\text{rel}}$)
based on the reliable \emph{a posteriori} information in the output ball radius.

Use of $\varepsilon_{\text{rel}}$
further depends circularly on $V$ ($V$ is essentially what we are trying
to compute!), so the algorithm must guess $V$\!.
A too large guess means loss of accuracy and a too small guess means
unnecessary work.
Our approach is to start with the tolerance
$\varepsilon_{\text{abs}}$
and continuously update $\varepsilon_{\text{abs}} \gets \max(\varepsilon_{\text{abs}}, I_a \varepsilon_{\text{rel}})$,
$\varepsilon_{\text{abs}} \gets \max(\varepsilon_{\text{abs}}, J_a \varepsilon_{\text{rel}})$
where $|I| = [I_a,I_b]$ and $|J| = [J_a,J_b]$ are intervals computed in steps 2 and 3;
$I_a$ and $J_a$ will then be lower bounds for $V$ (we err on the side of
preserving accuracy), modulo global cancellation in the integral.
As noted above, the user should exploit knowledge about $V$
if possible
since $I_a$ and $J_a$ may be pessimistic.
More clever globally adaptive strategies are possible,
but we settled for this simple approach in
the present version.

To abort gracefully when convergence
is too slow,
\emph{evaluation limits} include a bound on the number of
calls to $f$ (default $\smash{1000 p + p^2}$) and a bound
on the size~$N$ of the work queue $Q$ (default $2p$).
By default, $Q$ acts as a stack and step~4 puts the new subinterval
with the larger error at the top; optionally,
$Q$ can be switched to a global priority queue, which may
improve results if convergence is so slow
that evaluation limits are exceeded. This has the downside of
sometimes requiring $N$ nearly as large as the number of calls
to $f$ (e.g.\ $N \sim p^2$), whereas we always have $N \lesssim p$ with the stack.
A more clever algorithm might use a top-level priority queue
down to some depth before switching to a stack locally.

\begin{figure}[t]
\begin{centering}
\includegraphics[width=0.95\textwidth]{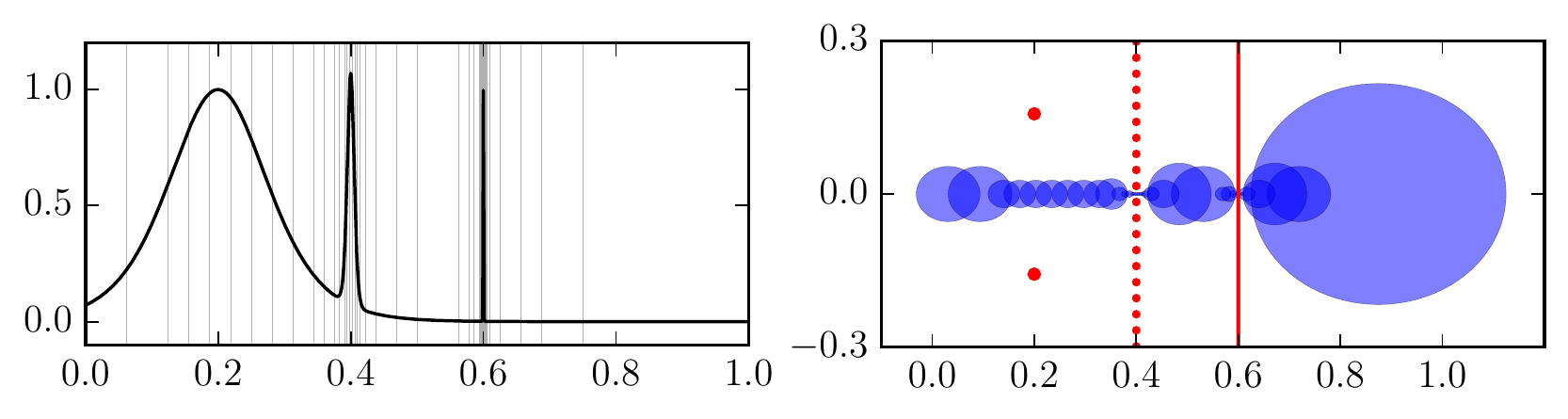}
\caption{Left: $f(x) = \operatorname{sech}^2({\scriptstyle 10(x-0.2)}) +
\operatorname{sech}^4({\scriptstyle 100(x-0.4)}) +
\operatorname{sech}^6({\scriptstyle 1000(x-0.6)})$, with subintervals used by Arb.
Right: complex ellipses used. The dots show the poles of $f$.}
\label{fig:spikeintegral}
\end{centering}
\vspace{-3mm}
\end{figure}

\vspace{-3mm}

\section{Benchmarks}

\vspace{-1mm}

We test various integrals
with precision between about 10 and 1000 digits.
Timings were obtained on an Intel Core i5-4300U CPU.
We compare Arb to the heuristic arbitrary-precision integration
routines \texttt{intnum} in Pari/GP~\cite{PARI2} and \texttt{quad} in \mbox{mpmath}~\cite{mpmath}.
Both use double exponential quadrature without adaptive subdivision,
although \texttt{quad} is degree-adaptive.
Further comparisons
with other numerical and interval packages
(as well as alternative methods in Pari/GP and mpmath\footnote{For example,
mpmath provides \texttt{quadgl} for Gaussian quadrature, which is 2-3 times
faster on some examples, but its precomputations are prohibitive at high precision.}) would be useful,
but out of scope for this brief overview.

We do not show the outputs, but note that in all cases,
Arb computes correct balls with radius a small multiple of $2^{-p}$.
On some test cases, mpmath with default settings silently returns an inaccurate answer
due to exceeding its limit on the quadrature degree,
but it provides an optional mechanism to catch this.
We increased the degree limit to let mpmath run to full accuracy in all cases,
and have written (!) after a timing where the default
is insufficient.
Pari/GP is not adaptive and silently returns inaccurate answers
without providing a catch mechanism or a way to increase the degree.
It does provide an option to split the interval
non-adaptively into $2^t$ parts, but it is up to the user
to find a correct~$t$. We have done so where necessary, which is also marked in the timings.

\pagebreak

\subsection{Integrals without singularities on the path}
\vspace{-1mm}

Table~\ref{tab:integrationtabanalytic}
shows examples with smooth $f$ on $[a,b]$.
For meromorphic~$f$, the number of subintervals largely depends
on the location of the poles and does not change with $p$.
The ``spike integral'' $I_1$ (Figure~\ref{fig:spikeintegral})
is a well known pathological example~\cite{cranley1971,hale2010}; all ordinary
numerical integrators we have tested
(Mathematica, GSL, SciPy, etc.) give inaccurate results with default settings.
This integrand has poles near the real axis, forcing many local
bisections. It is a piece of cake for the Petras algorithm,
but Pari/GP and mpmath converge slowly
unless the user manually splits the path at the peaks.
$I_2$ could be sped up 40\% in Arb by using
$\cos^2(x) = \tfrac{1}{2}(1\!+\!\cos(2x))$ for wide~$x$
to bound the denominator more tightly.

For entire functions ($I_4, I_5, I_6$), the efficiency improves
with larger $p$ since arbitrarily large bounding ellipses can be used.
$I_5$ is Rump's example again, and~$I_6$ (whose graph has two sharp ``bends'')
was provided by Silviu-Ioan Filip.

The code is seen to work well with special functions.
In $I_3$ we integrate the Lambert $W$ function, where we need
to check for the branch cut on $(-\infty,-1/e]$ in the evaluation.
$I_7$ also illustrates integration on a complex path.

Overall, Arb is faster than Pari/GP and mpmath,
despite the fact that rigorous error bounds
create extra work.
The speedup is in part explained by faster arithmetic and
transcendental functions in Arb and lower overhead due to using~C,
as well as the advantage of Gaussian quadrature over the double
exponential method for smooth integrands.
However, if these differences are accounted for,
we can still conclude that the Petras algorithm in
ball arithmetic holds up extremely well for high-precision
integration, on top of giving rigorous bounds.

\begin{table}[t!]
\renewcommand{\arraystretch}{0.9}
\setlength{\tabcolsep}{.3em}
\begin{center}
\caption{Integrals without singularities on $[a,b]$. Timings
(Pari/GP, mpmath, Arb) are in seconds.
Sub = number of terminal subintervals (requiring no further bisection) used by Arb,
Eval = total number of integrand evaluations used by Arb.}
\resizebox{0.9 \width}{!} {
\begin{tabular}{|r | l l l r r | l l l r r |}
       \hline
       $p$ & Pari/GP & mpmath & Arb & \!\!Sub & Eval
           & Pari/GP & mpmath & Arb & \!\!Sub & Eval \\ \hline

       & \multicolumn{5}{|c|}{$\phantom{\!\!\!\!\!\!\!\!\!\!\!\!\!\left|e^{x^2}\right|}
            I_0 = \int_0^1 1/(1+x^2) dx$}
       & \multicolumn{5}{|c|}{$\phantom{\!\!\!\!\!\!\!\!\!\!\!\!\!\left|e^{x^2}\right|}
            I_1 = \int_0^1 \sum_{k=1}^3 \operatorname{sech}^{2k}(10^k(x-0.2k)) \, dx$}

        \\

       32 & 0.00039 & 0.00057 & 0.000025\!\! & 2 & 32    & 0.54 {\scriptsize\!($t\!=\!8$)}\! & 1.9 {\scriptsize\!(!)} &   0.0030 & 49 & 795  \\
       64 & 0.00039 & 0.0011 & 0.000036\!\! & 2 & 52     & 0.54 {\scriptsize\!($t\!=\!8$)}\! & 5.0 {\scriptsize\!(!)} &   0.0051 & 49 & 1299 \\
       333 & 0.0043 & 0.0058 & 0.00018 & 2 & 188         & 12 {\scriptsize\!($t\!=\!9$)}\! & 38 {\scriptsize\!(!)} &   0.038 & 49 & 4891 \\
       3333 & 1.0 & 0.13 & 0.014 & 2 & 2056              & 3385 {\scriptsize\!($t\!=\!9$)}\! & -  &   8.7 & 49 & 48907 \\ \hline

       & \multicolumn{5}{|c|}{$\phantom{\!\!\!\!\!\!\!\!\!\!\!\!\!\left|e^{x^2}\right|}
            I_2 = \int_0^{\pi} x \sin(x) / (1\!+\!\cos^2(x)) dx$}
       & \multicolumn{5}{|c|}{$\phantom{\!\!\!\!\!\!\!\!\!\!\!\!\!\left|e^{x^2}\right|}
            I_3 = \int_0^{1000} W_0(x) dx$}
        \\

       32 & 0.00077 & 0.0021 & 0.00033\! & 14 & 229    & 0.0037 & 0.012 &   0.00041 & 12 & 163  \\
       64 & 0.00077 & 0.0046 & 0.00054\! & 14 & 373    & 0.0037 & 0.032 &   0.00093 & 12 & 273 \\
       333 & 0.0088 & 0.037 & 0.0040 & 14 & 1401       & 0.052 {\scriptsize\!($t\!=\!1$)}\! & 0.25  &   0.0099 & 12 & 1109 \\
       3333 & 2.2 & 4.4 & 1.0 & 14 & 14401             & 11 {\scriptsize\!($t\!=\!2$)}\! & 25 &   1.3 & 12 & 12043 \\ \hline

       & \multicolumn{5}{|c|}{$\phantom{\!\!\!\!\!\!\!\!\!\!\!\!\!\left|e^{x^2}\right|}
            I_4 = \int_0^{100} \sin(x) dx$}
       & \multicolumn{5}{|c|}{$\phantom{\!\!\!\!\!\!\!\!\!\!\!\!\!\left|e^{x^2}\right|}
            I_5 = \int_0^{8} \,\sin\!\left(x+e^x\right) dx$}
        \\

       32 & 0.0012 {\scriptsize\!($t\!=\!1$)}\! & 0.0019 & 0.000047 & 1 & 53    & 0.063 {\scriptsize\!($t\!=\!6$)}\! & 0.23 {\scriptsize\!(!)} &   0.0048 & 33 & 2115   \\
       64 & 0.0012 {\scriptsize\!($t\!=\!1$)}\! & 0.0014 & 0.000074 & 1 & 72    & 0.063 {\scriptsize\!($t\!=\!6$)}\! & 0.25 {\scriptsize\!(!)} &   0.0055 & 27 & 2307 \\
       333 & 0.015 {\scriptsize\!($t\!=\!1$)}\! & 0.018 & 0.00030 & 1 & 139     & 0.22 {\scriptsize\!($t\!=\!4$)}\! & 0.58 {\scriptsize\!(!)} &   0.017 & 22 & 4028 \\
       3333 & 2.0 & 0.71 & 0.032 & 1 & 526                                      & 14 {\scriptsize\!($t\!=\!2$)}\! & 12 &                          1.1 & 8 & 10417 \\ \hline

       & \multicolumn{5}{|c|}{$\phantom{\!\!\!\!\!\!\!\!\!\!\!\!\!\left|e^{x^2}\right|}
            I_6 = \int_{-1}^{1} e^{-x} \operatorname{erf}\!\left(\!\sqrt{1250} \, x + \tfrac{3}{2}\right) dx$}
       & \multicolumn{5}{|c|}{$\phantom{\!\!\!\!\!\!\!\!\!\!\!\!\!\left|e^{x^2}\right|}
            I_7 = \int_1^{1+1000i} {\mathrm \Gamma(x)} dx$}
        \\

       32 & 0.024 {\scriptsize\!($t\!=\!3$)}\! & 0.018 {\scriptsize\!(!)} & 0.0025 & 7 & 297   & 0.031 {\scriptsize\!($t\!=\!2$)}\! & 0.028 &   0.00076 & 11 & 103   \\
       64 & 0.024 {\scriptsize\!($t\!=\!3$)}\! & 0.057 {\scriptsize\!(!)} & 0.0055 & 6 & 438   & 0.054 {\scriptsize\!($t\!=\!3$)}\! & 0.093 &   0.0035 & 12 & 280  \\
       333 & 0.50 {\scriptsize\!($t\!=\!3$)}\! & 0.22 & 0.047 & 4 & 791                        & 0.65 {\scriptsize\!($t\!=\!3$)}\! & 1.1 &   0.081 & 14 & 1304 \\
       3333 & 173 {\scriptsize\!($t\!=\!2$)}\! & 466 & 5.7 & 2 & 2923                          & 561 {\scriptsize\!($t\!=\!3$)}\! & 847 &   48 & 14 & 16535 \\ \hline

\end{tabular}
}
\label{tab:integrationtabanalytic}
\vspace{-8mm}
\end{center}
\end{table}



\pagebreak

\begin{table}[t!]
\renewcommand{\arraystretch}{0.9}
\setlength{\tabcolsep}{.3em}
\begin{center}
\caption{Improper integrals and integrals with endpoint singularities.
For integration with Arb, all improper integrals (i.e.\ excluding $E_0$)
have been truncated manually at a lower bound $\varepsilon$
or upper bound $N$, chosen so that the omitted part is smaller than $2^{-p}$.}
\resizebox{0.9 \width}{!} {
\begin{tabular}{|r | l l l r r | l l l r r |}
       \hline
       $p$ & Pari/GP & mpmath & Arb & \!\!Sub & Eval
           & Pari/GP & mpmath & Arb & \!\!Sub & Eval \\ \hline

       & \multicolumn{5}{|c|}{$\phantom{\!\!\!\!\!\!\!\!\!\!\!\!\!\left|e^{x^2}\right|}
            E_0 = \int_0^1 \sqrt{1-x^2} dx$}
       & \multicolumn{5}{|c|}{$\phantom{\!\!\!\!\!\!\!\!\!\!\!\!\!\left|e^{x^2}\right|}
            E_1 = \int_0^{\infty} 1/(1+x^2) \, dx$}

        \\

       32 & 0.00041 & 0.00055 & 0.00022 & \!\!\!\!22 & 234     & 0.00060 & 0.0010 &   0.00079 & \!\!\!\!94 & 997   \\
       64 & 0.00041 & 0.00067 & 0.00057 & \!\!\!\!44 & 674      & 0.00060 & 0.0012 &   0.0022 & \!\!\!\!190 & 2887  \\
       333 & 0.0044 & 0.0060 & 0.015 & \!\!\!\!223 & 12687     & 0.0068 & 0.011 &   0.048 & \!\!\!\!997 & 51900  \\
       3333 & 0.94 & 0.18 & 6.6 & \!\!\!\!2223 & 1187293        & 1.7 & 0.24 &   27 & \!\!\!\!9997 & 4711128  \\ \hline

       & \multicolumn{5}{|c|}{$\phantom{\!\!\!\!\!\!\!\!\!\!\!\!\!\left|e^{x^2}\right|}
            E_2 = \int_0^1 \log(x) / (1+x) dx$}
       & \multicolumn{5}{|c|}{$\phantom{\!\!\!\!\!\!\!\!\!\!\!\!\!\left|e^{x^2}\right|}
            E_3 = \int_0^{\infty} \operatorname{sech}(x) \, dx$}

        \\

       32 & 0.00081 & 0.00080 & 0.00042 & \!\!\!\!34 & 361        & 0.0011 & 0.0019 &   0.00017 & 9 & 144   \\
       64 & 0.00081 & 0.00094 & 0.0012 & \!\!\!\!67  & 1026      & 0.0011 & 0.0043 &   0.00032 & 10 & 251  \\
       333 & 0.011 & 0.011 & 0.038  & \!\!\!\!336 & 19254         & 0.013 & 0.098 &   0.0030 & 14 & 1277  \\
       3333 & 1.7 & 1.08 & 106 & \!\!\!\!3336 & 1787191          & 3.5 & 3.3 &   0.95 & 17 & 16593  \\ \hline

       & \multicolumn{5}{|c|}{$\phantom{\!\!\!\!\!\!\!\!\!\!\!\!\!\left|e^{x^2}\right|}
            E_4 = \int_0^{\infty} e^{-x^2+ix} dx$}
       & \multicolumn{5}{|c|}{$\phantom{\!\!\!\!\!\!\!\!\!\!\!\!\!\left|e^{x^2}\right|}
            E_5 = \int_0^{\infty} e^{-x} \operatorname{Ai}(-x) \, dx$}

        \\

       32 & 0.0014 & 0.0067 & 0.00011 & 1 & 71       & - & 0.19   & 0.0028 & 4 & 269   \\
       64 & 0.0014 & 0.016  & 0.00018 & 1 & 98       & - & 0.91 {\scriptsize\!(!)}   & 0.012 & 9 & 842  \\
       333 & 0.017 & 0.13   & 0.0016 & 2 & 397       & - & 26 {\scriptsize\!(!)}   & 0.94 & 124 & 24548  \\
       3333 & 4.7 & 7.1     & 0.47 & 4 & 3894        & - & 10167 {\scriptsize\!(!)}   & 502 & 1205 & 709889  \\ \hline

\end{tabular}
}
\label{tab:integrationtabsingular}
\end{center}
\vspace{-7mm}
\end{table}


\subsection{Endpoint singularities and infinite intervals}
\vspace{-2mm}

The methods in Pari/GP and mpmath are designed to
support typical integrals with infinite intervals or endpoint singularities,
which often arise in applications.
Arb requires finite $a,b$ and a bounded $f$ to return a finite result, but
the user may provide a manual truncation
(say $\smash{\int_0^{\infty}\!\!f(x) dx \approx \int_{\varepsilon}^N\!\!f(x) dx}$)
to work around this restriction.
Tail bounds must then be added based on symbolic knowledge about~$f$.
This is not ideal in terms of usability or efficiency,
but since the Petras algorithm works well even
with an endpoint very close to a singularity (or~$\infty$),
evaluating improper integrals to high precision in this way is at least feasible.\footnote{An exception
is when $f$ has an essential singularity
inducing oscillation combined with slow decay. Oscillation with
exponential decay is not a problem (as in~$E_4$,~$E_5$), but integrals
like $\smash{\int_0^1\!\sin(1/x) dx \!=\! \int_1^{\infty} \!\sin(x)/x^2}$ (not benchmarked here)
require $\smash{2^{O(p)}\!}$ work, so we can only hope for 5-10 digits without specialized oscillatory algorithms.}

In Table~\ref{tab:integrationtabsingular},
$E_0$, $E_1$ and $E_2$ have algebraic or logarithmic singularities
or decay, with $E_1$ requiring $N \approx 2^p$ and $E_2$ requiring $\varepsilon \approx 2^{-p}$
(no truncation is needed for $E_0$, as $f$ is bounded at the algebraic branch point singularity $x = 1$).
Here Arb needs $O(p)$ subintervals and $O(p^2)$ evaluations,
while the double exponential algorithm in Pari/GP and mpmath only needs roughly $O(p)$ evaluations
and therefore scales better.\footnote{As a means
to improve performance, we note the standard trick of manually changing variables to turn
algebraic growth or decay into exponential decay.
Indeed, $x \to \sinh(x)$ gives $E_1 = E_3$.
Similarly $x \to \tanh(x)$ and $\smash{x \to e^{-x}}$ can be used in $E_0$, $E_2$.}
For integrals with exponential decay ($E_3$, $E_4$ and~$E_5$),
a cutoff of $N \sim p$ is sufficient,
and here Arb retains excellent performance.

In a future extension of this work,
some reasonable class of improper integrals
could be supported more efficiently
and conveniently (e.g. with the user providing
a symbolic bound like $|f(x)| < C x^{\alpha} \exp(-\beta x^{\gamma})$).


\pagebreak

\subsection{Piecewise and discontinuous functions}

Piecewise real analytic functions can be integrated
efficiently using piecewise complex analytic extensions.
For example, $|x|$ on $\mathbb{R}$ extends to the
function $\sqrt{z^2}$ of $z=x+yi$,
which equals $z$ in the right plane and $-z$ in the left plane
with a branch cut on
$\operatorname{Re}(z) = 0$.\footnote{This works for integrating $|f|$ when $f$ is real,
but since $|\cdot|$ on $\mathbb{C}$ is not holomorphic,
integrating $|f|$ for nonreal $f$ must use direct enclosures, with $2^{O(p)}$ cost.
In that case, the user should instead construct
complex-extensible real and imaginary parts
$f = g\!+\!h i$ (e.g.\ via Taylor polynomials if no closed forms exist)
and integrate $\sqrt{g^2 + h^2}$.}
We provide as library methods such extensions of
$\operatorname{sgn}(x)$, $|x|$, $\lfloor x \rfloor$, $\lceil x \rceil$,
$\max(x,y)$, $\min(x,y)$, with builtin branch cut detection.

\begin{table}[t!]
\renewcommand{\arraystretch}{0.9}
\setlength{\tabcolsep}{.3em}
\begin{center}
\caption{Integrals with point discontinuities in $f$ or $f'$. Here $p(x) = x^4 + 10x^3 + 19x^2 - 6x - 6$ in $D_0$,
and $u(x) = (x\!-\!\lfloor x \rfloor\!-\!\tfrac{1}{2})$, $v = \max(\sin(x),\cos(x))$ in $D_3$. For $D_3$,
the function evaluation limit had to be increased for convergence at $p = 3333$.}
\resizebox{0.9 \width}{!} {
\begin{tabular}{|r | l r r | l r r | l r r | l r r |}
       \hline
       $p$ & Arb & \!\!Sub & Eval
           & Arb & \!\!Sub & Eval
           & Arb & \!\!Sub & Eval
           & Arb & \!\!Sub & Eval \\ \hline

       & \multicolumn{3}{|c|}{$\phantom{\!\!\!\!\!\!\!\!\!\!\!\!\!\left|e^{x^2}\right|}
            D_0 = \int_0^1 |p(x)| \, e^x \, dx$}
       & \multicolumn{3}{|c|}{$\phantom{\!\!\!\!\!\!\!\!\!\!\!\!\!\left|e^{x^2}\right|}
            D_1 = \int_0^{100} \lceil x \rceil \, dx$}
       & \multicolumn{3}{|c|}{$\phantom{\!\!\!\!\!\!\!\!\!\!\!\!\!\left|e^{x^2}\right|}
            D_2 = \int_{-1-i}^{-1+i} \sqrt{x} \, dx$}
       & \multicolumn{3}{|c|}{$\phantom{\!\!\!\!\!\!\!\!\!\!\!\!\!\left|e^{x^2}\right|}
            D_3 \!=\! \int_0^{10} \! u(x) v(x) dx$}
        \\

       32 & 0.00058\!\!\! & \!38 & 412     &   0.0054 & \!\!\!2208 & 6622       & 0.00064\!\!\! & 68 & 506     &   0.011 & \!\!699 & \!\!5891   \\
       64 & 0.0016\!\! & \!70 & 1093     &   0.014 & \!\!\!5536 & 16606         & 0.0021\!\! & 132 & 1462     &   0.035 & \!\!1437 & \!\!\!\!19653  \\
       333 & 0.049\! & \!\!339 & \!18137        &   0.12 & \!\!\!\!\!33512 & 100534     & 0.067\! & \!670 & 28304        &   1.4 & \!\!7576 & \!\!\!\!436 K  \\
       3333 & 101 & \!\!\!3339 & \!1624951        &   1.6 & \!\!\!\!\!\!345512 & 1036534    & 35 & \!\!6670 & 2669940        &   2805 & \!\!76101 & \!\!\!42 M  \\ \hline

\end{tabular}
}
\label{tab:integrationtabdisc}
\end{center}
\vspace{-7mm}
\end{table}

Table~\ref{tab:integrationtabdisc} shows integrals with mid-interval jumps
or kinks, including one complex integral crossing a branch cut discontinuity ($D_2$).
The example $D_0$, where~$p(x)$
changes sign once on $[0,1]$,
is due to Helfgott (see comments in~\cite{mahboubi2016formally}).

We see that a mid-interval singularity
leads to use of $O(p)$ subintervals and $O(p^2)$ evaluations to isolate
the problematic point by bisection.
With~$k$ such points ($D_1$ and $D_3$), the cost simply increases
by another factor $k$, and the user may have to raise the evaluation limits accordingly
to let the algorithm complete
(which we did for $D_3$).
In contrast, Pari/GP and mpmath
cope poorly with mid-interval singularities
and cannot achieve high accuracy on these examples unless the user
manually splits
the interval precisely at the problematic points.

\vspace{-1mm}
\section{Complex analysis}
\vspace{-1mm}

We conclude by illustrating integration as a tool for complex analysis.
First, we consider computing derivatives via the Cauchy integral formula.
Denote by $\smash{\wp(z; \tau) \!=\! \sum_{n=-2}^{\infty} \!a_n(\tau) z^n}$ the
Weierstrass elliptic function for the lattice $(1,\tau)$.
We fix $\tau = i$ (placing the poles of $\wp$ at the
Gaussian integers) and compute the Laurent coefficients
$a_n = \smash{\frac{1}{2\pi i} \int_\gamma z^{-n-1} \wp(z) dz}$ by
integrating along the square connecting $\pm 0.5 \pm 0.5i$.
We ignore symmetry and compute all four segments.
With $p = 333$, some results are
(note that $a_{-1} = a_{100} = 0$ and all $a_n$ are real):

\begin{small}
\begin{verbatim}
a[-2]  = [1.00000000000000000000000... +/- 3.57e-98] + [+/- 1.89e-98]*I
a[-1]  =                              [+/- 4.11e-98] + [+/- 2.57e-98]*I
a[2]   = [9.45363600646169261465306... +/- 4.44e-97] + [+/- 2.48e-97]*I
a[98]  = [395.999999999999648281345... +/- 2.90e-68] + [+/- 1.17e-68]*I
a[100] =                              [+/- 4.95e-68] + [+/- 4.95e-68]*I
\end{verbatim}
\end{small}

\pagebreak

We lose about $n$ bits of precision to cancellation due to the
integrand magnitude growing with $n$.
Apart from this, the difficulty increases quite slowly with~$n$: $a_{-2}$ takes 0.67 s
while $a_{98}$ and $a_{100}$ take 0.85 s at this precision.

As a second example, the number $N(T)$ of zeros $\rho_k$ of the
Riemann zeta function $\zeta(s)$ on the box $[0,1] + [0,T] i$ can
be computed via the argument principle
$$N(T)-1 = \frac{1}{2 \pi i}\! \int_{\gamma} \frac{\zeta'(s)}{\zeta(s)} ds =
\frac{\theta(T)}{\pi} + \frac{1}{\pi} \operatorname{Im}\!\! \left[
\int_{1+\varepsilon}^{1+\varepsilon+T i} \! \frac{\zeta'(s)}{\zeta(s)} ds +
\!\int_{1+\varepsilon+T i}^{\tfrac{1}{2} + T i} \! \frac{\zeta'(s)}{\zeta(s)} ds \right]$$
where $\gamma$ traces the boundary of $[-\varepsilon, 1+\varepsilon] + [0,T] i$
(plus an excursion for the pole at $s=1$, whence the $-1$ term).
The more numerically useful formula on the right, where $\varepsilon > 0$
now is arbitrary, is a well-known consequence
of the functional equation, where $\theta(T)$ is the Hardy theta function.
We set $\varepsilon = 99$ (!) so that only the horizontal segment is difficult,
and evaluate the integrals with $\varepsilon_{\text{abs}} = 10^{-6}$:
\vspace{-1mm}

\begin{center}
\renewcommand{\arraystretch}{0.7}
\setlength{\tabcolsep}{.4em}
\resizebox{0.9 \width}{!} {
\begin{tabular}{l l l r r l}
$T$ & $p$ & Time (s) & Sub & Eval & $N(T)$ \\
\hline\rule{0pt}{2.2ex}$10^3$ & $32$ & 0.51 & 109 & 1219 & {\small \texttt{[649.00000 +/- 7.78e-6]}} \\
$10^5$ & $32$ & 12 & 353 & 4088 & {\small \texttt{[138069.000 +/- 3.10e-4]}} \\
$10^7$ & $48$ & 42 & 391 & 4500 & {\small \texttt{[21136125.0000 +/- 5.53e-5]}} \\
$10^9$ & $48$ & 1590 & 677 & 8070 & {\small \texttt{[2846548032.000 +/- 1.95e-4]}} \\
\end{tabular}
}
\end{center}

\vspace{-1mm}

We obtain balls that provably determine $N(T)$, and
the method scales reasonably well.
Unfortunately, the evaluation of $\zeta(s)$ in Arb is
currently not well tuned for all $s$, which
makes large~$T$ slower than necessary
and can make this computation extremely slow with slightly different settings.
In general, for complicated integrals, the user may need to customize the integrand evaluation
to handle wide balls
or large parameters optimally for a given path and precision.

\bibliographystyle{plain}
\bibliography{references}

\begin{thebibliography}{10}

\bibitem{bailey2011high}
D.~H. Bailey and J.~M. Borwein.
\newblock High-precision numerical integration: Progress and challenges.
\newblock {\em Journal of Symbolic Computation}, 46(7):741--754, 2011.

\bibitem{cranley1971}
R.~Cranley and T.~N.~L. Patterson.
\newblock On the automatic numerical evaluation of definite integrals.
\newblock {\em The Computer Journal}, 14(2):189--198, 1971.

\bibitem{hale2010}
N.~Hale.
\newblock Spike integral.
\newblock {\scriptsize
  \url{http://www.chebfun.org/examples/quad/SpikeIntegral.html}}, 2010.

\bibitem{Johansson2017arb}
F.~Johansson.
\newblock Arb: efficient arbitrary-precision midpoint-radius interval
  arithmetic.
\newblock {\em IEEE Transactions on Computers}, 66:1281--1292, 2017.

\bibitem{mpmath}
F.~Johansson.
\newblock mpmath version 1.0.
\newblock {\scriptsize \url{http://mpmath.org/}}, 2017.

\bibitem{JohanssonMezzarobba2018fast}
F.~Johansson and M.~Mezzarobba.
\newblock Fast and rigorous arbitrary-precision computation of
  {G}auss-{L}egendre quadrature nodes and weights.
\newblock {\em arXiv:1802.03948}, 2018.

\bibitem{mahboubi2016formally}
A.~Mahboubi, G.~Melquiond, and T.~Sibut-Pinote.
\newblock Formally verified approximations of definite integrals.
\newblock In {\em International Conference on Interactive Theorem Proving},
  pages 274--289. Springer, 2016.

\bibitem{petras2002self}
K.~Petras.
\newblock Self-validating integration and approximation of piecewise analytic
  functions.
\newblock {\em J. Comp. Appl. Math.}, 145(2):345--359, 2002.

\bibitem{rump2010verification}
S.~M. Rump.
\newblock Verification methods: Rigorous results using floating-point
  arithmetic.
\newblock {\em Acta Numerica}, 19:287--449, 2010.

\bibitem{PARI2}
{The Pari group}.
\newblock {Pari/GP version 2.9.4}.
\newblock {\scriptsize \url{http://pari.math.u-bordeaux.fr/}}, 2017.

\bibitem{sage}
{The SageMath developers}.
\newblock {SageMath}.
\newblock {\scriptsize \url{http://sagemath.org/}}, 2005--.

\end{thebibliography}

\end{document}